\documentclass[aps,prl, amsmath, showpacs, preprintnumbers,superscriptaddress, twocolumn,sort&compress,floatfix, amssymb]{revtex4}
\pdfoutput=1
\usepackage{graphicx}
\usepackage{rotating}
\usepackage{dcolumn}
\usepackage{bm}
\usepackage{color}
\usepackage{mathptmx, textcomp}
\usepackage[latin1]{inputenc}
\usepackage{braket}

\usepackage{multirow}
\newcommand{\minitab}[2][l]{\begin{tabular}{#1}#2\end{tabular}}

\begin{document}

\author{F. Meinert}
\affiliation{Institut f\"ur Experimentalphysik und Zentrum f\"ur Quantenphysik, Universit\"at Innsbruck, 6020 Innsbruck, Austria}
\author{M. Panfil}
\affiliation{SISSA-International School for Advanced Studies and INFN, Sezione di Trieste, 34136 Trieste, Italy}
\author{M. J. Mark}
\affiliation{Institut f\"ur Experimentalphysik und Zentrum f\"ur Quantenphysik, Universit\"at Innsbruck, 6020 Innsbruck, Austria}
\affiliation{Institut f\"ur Quantenoptik und Quanteninformation, \"Osterreichische Akademie der Wissenschaften, 6020 Innsbruck, Austria}
\author{K. Lauber}
\affiliation{Institut f\"ur Experimentalphysik und Zentrum f\"ur Quantenphysik, Universit\"at Innsbruck, 6020 Innsbruck, Austria}
\author{J.-S. Caux}
\affiliation{Institute for Theoretical Physics, University of Amsterdam, 1090 GL Amsterdam, The Netherlands}
\author{H.-C. N\"agerl}
\affiliation{Institut f\"ur Experimentalphysik und Zentrum f\"ur Quantenphysik, Universit\"at Innsbruck, 6020 Innsbruck, Austria}

\title{Probing the Excitations of a Lieb-Liniger Gas from Weak to Strong Coupling}

\date{\today}


\begin{abstract}
We probe the excitation spectrum of an ultracold one-dimensional Bose gas of Cesium atoms with repulsive contact interaction that we tune from the weakly to the strongly interacting regime via a magnetic Feshbach resonance. The dynamical structure factor, experimentally obtained using Bragg spectroscopy, is compared to integrability-based calculations valid at arbitrary interactions and finite temperatures. Our results unequivocally underly the fact that hole-like excitations, which have no counterpart in higher dimensions, actively shape the dynamical response of the gas.
\end{abstract}

\maketitle

Interacting quantum systems confined to a one-dimensional (1D) geometry display qualitatively different behavior compared to their higher-dimensional counterparts \cite{Giamarchi04}. 
Systems of strongly interacting electrons recently realized in electronic nanostructure devices \cite{Auslaender02,Barak10} have evidenced the breakdown of Landau's Fermi liquid theory of quasi-particles in 1D, a world in which new types of excitations emerge out of the inevitably collective nature of the dynamics. The understanding of these requires approaches going beyond Landau's paradigm. The best-known, valid for sufficiently small temperatures and energies, is the Luttinger Liquid (LL) formalism \cite{Haldane81}. When probing dynamical correlation functions, one however typically leaves this low-energy and large-wavelength limit and enters a regime where even recent extensions of the LL formalism to higher energies \cite{Imambekov09,Imambekov12} cannot capture all features. Instead, one must rely on nonperturbative calculations to understand the correct basis of excitations and quantitatively explain experiments, a recent example being spinon dynamics in quantum spin chains \cite{2013_Mourigal_NATPHYS_9,2013_Lake_PRL_111}. These systems however lack the tunability required to track the whole transformation occurring between the limits of weak and strong coupling.

\begin{figure}
\includegraphics[width=0.57\columnwidth]{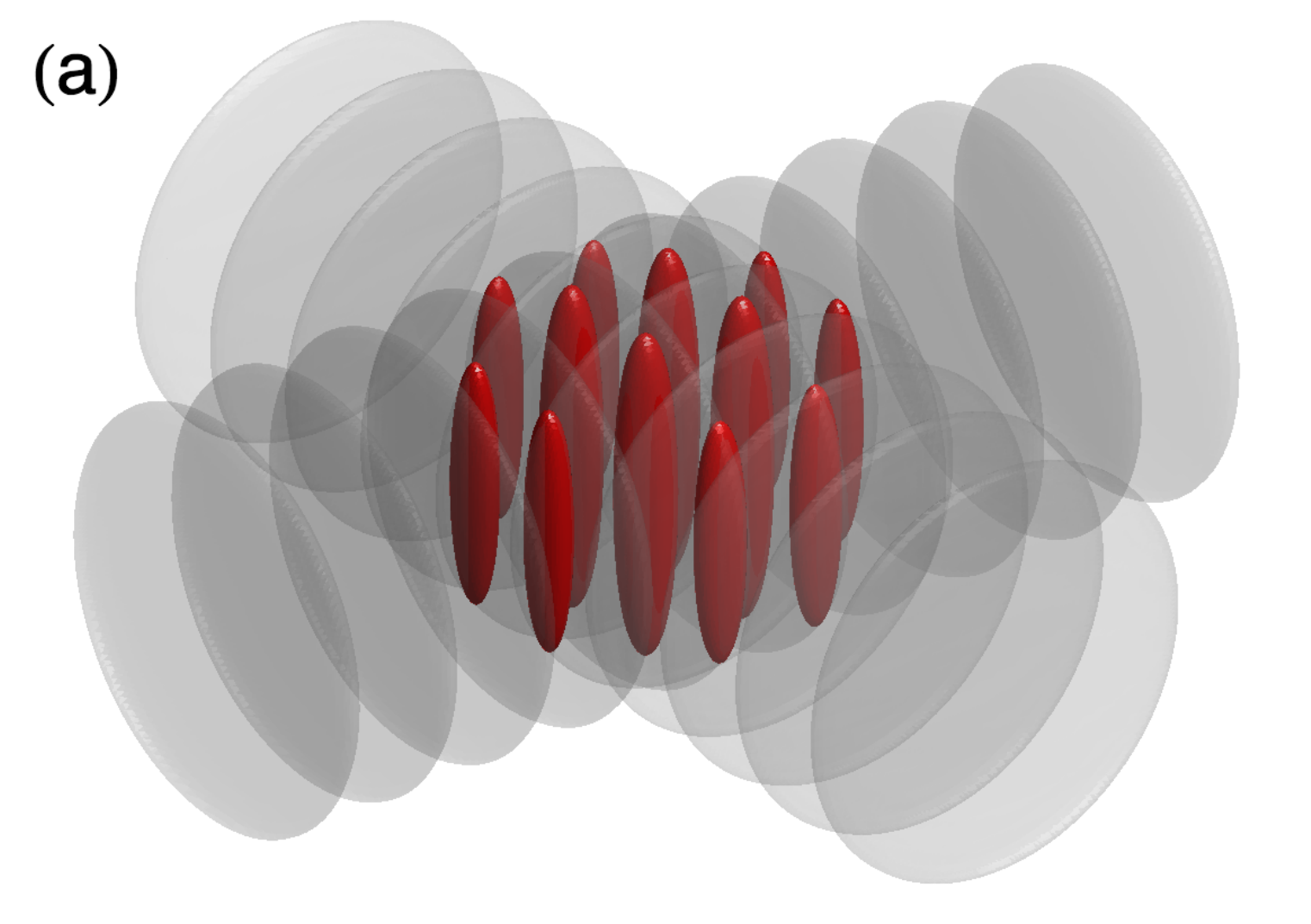}
\vspace{2mm}
\includegraphics[width=0.41\columnwidth]{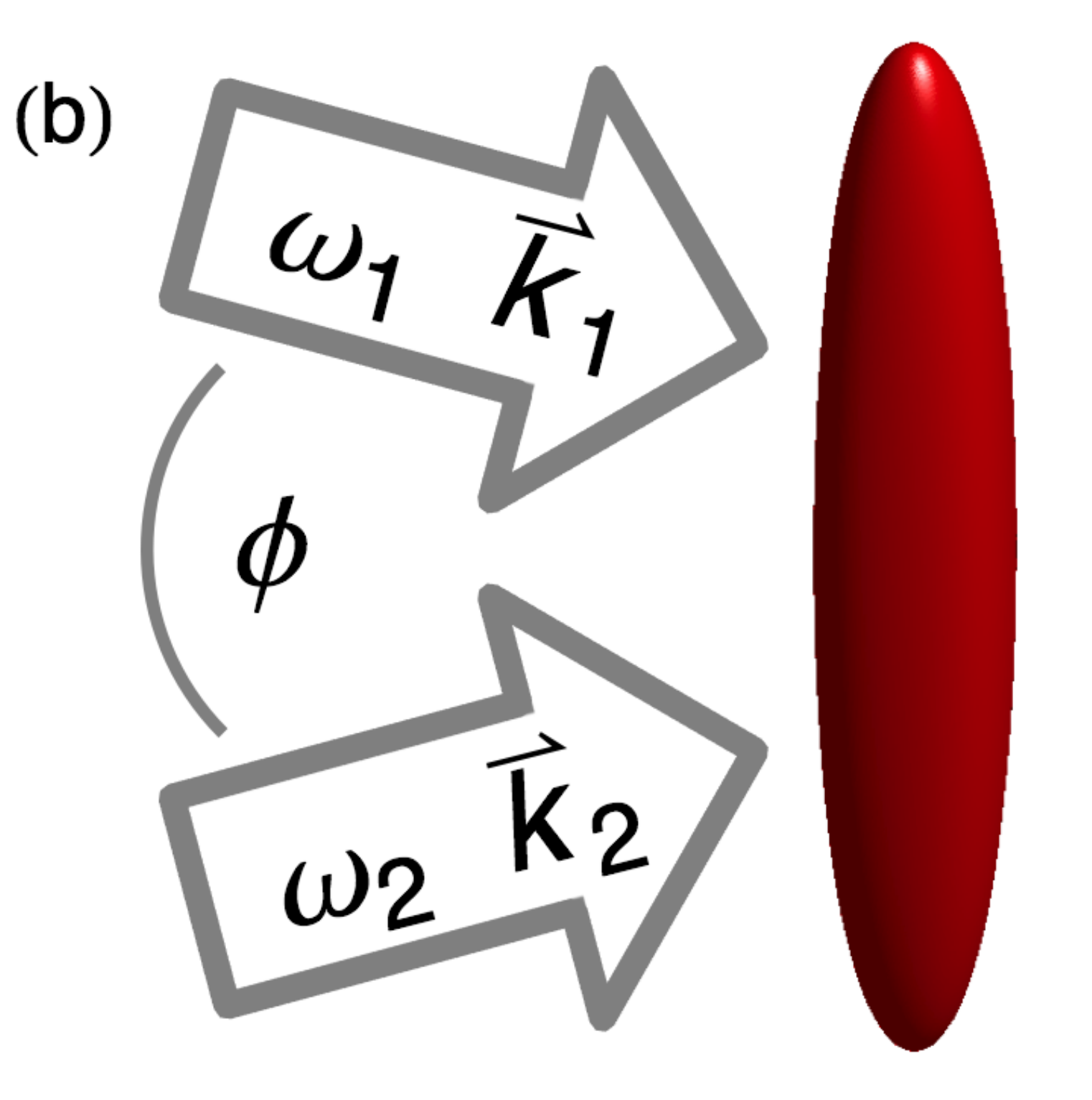}\\
\hspace{1mm}
\includegraphics[width=0.48\columnwidth]{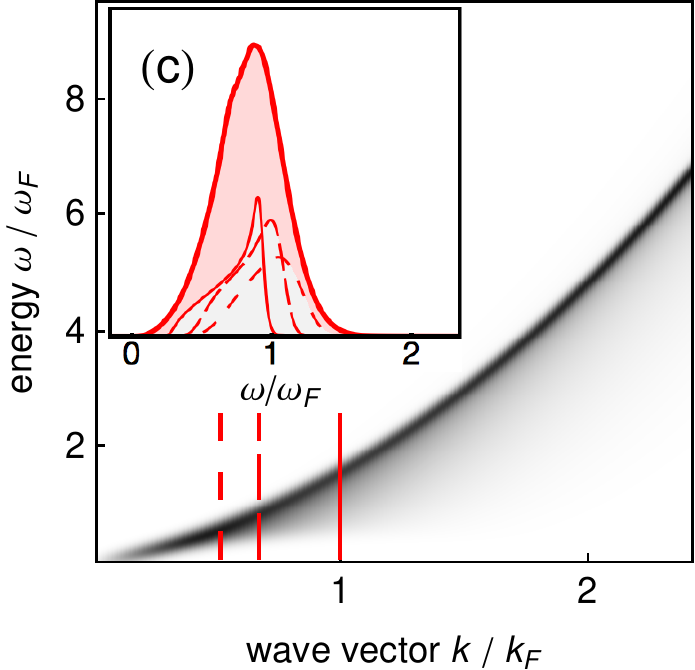}
\vspace{2mm}
\includegraphics[width=0.48\columnwidth]{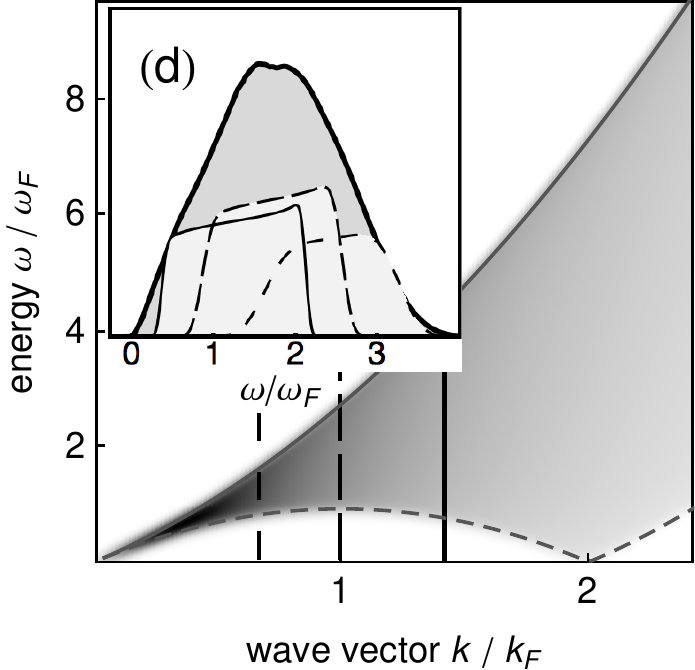}
\caption{\label{FIG1}(color online). Sketch of the experimental setup. (a) A pair of retro-reflected laser beams creates an ensemble of $\approx 4000$ independent one-dimensional Bose gases. (b) The excitation spectrum is probed by illuminating the gas with a pair of Bragg laser beams. (c),(d) Zero-temperature dynamical structure factor $S(k,\omega)$ (value shown in gray scale) for a moderately (c) ($\gamma=3.3$) and strongly (d) ($\gamma=45$) interacting homogeneous gas. The solid (dashed) line in (d) shows the dispersion of the Lieb-I (Lieb-II) mode. Insets indicate averaging over an ensemble of trapped systems. The thin lines show fixed momentum cuts at representative densities. The corresponding values for $k/k_F$ are indicated as vertical lines in the $k$-$\omega$-plane. The thick line shows the averaged response $S(\omega)$ in a local density approximation.}
\end{figure}

Very recently, systems of ultracold bosons have opened up new routes to study strong correlation effects in 1D \cite{Cazalilla11}, such as the fermionized Tonks-Girardeau (TG) gas of bosons \cite{Girardeau60,Haller09,Kinoshita04}, primarily due to unprecedented control over system parameters, e.g. confinement, particle interactions or quantum statistics \cite{Pagano14}. Moreover, the 1D Bose gas with contact interactions is one of the few integrable many-body problems that allows for combining experiment with numerically exact studies of the excitation spectrum, making it an important cornerstone on the path towards understanding interaction effects on dynamical correlation functions~\cite{Fabbri14}. In their seminal work \cite{Lieb63,Lieb63b}, Lieb and Liniger have shown that next to a particle-like mode (Lieb-I mode), which resembles Bogoliubov excitations in the limit of weak interactions, a second mode naturally emerges (Lieb-II mode) that stems from hole-like excitations in the effective Fermi sea in 1D. The coexistence of these two types of elementary excitations leads to a significant broadening of the dynamical response functions, clearly visible in the strongly interacting regime \cite{Caux06,Cherny06} (see Fig.~\ref{FIG1}).

In this Letter, we measure the dynamical structure factor (DSF) of the Lieb-Liniger Bose gas realized with ultracold atoms confined to 1D quantum tubes with widely tunable interactions. The analysis is based on careful disentangling of the experimental traits and allows us to identify the role of the Lieb-Liniger dynamics in shaping the response of the system. Comparison of the measured spectra with state-of-the-art numerical calculations \cite{Caux06,Panfil14} ranging from the weakly to the strongly interacting regime allows for a clear distinction between interaction and temperature effects and demonstrate the contribution of the Lieb-II type excitations to the response of the system.  

Our experiment starts with a Cesium Bose-Einstein condensate (BEC) of typically $1.1 \times 10^5$ atoms confined in a crossed dipole trap \cite{Weber2002,Kraemer2004}. The BEC is adiabatically loaded into an array of $\approx 4000$ quantum wires created via two mutually perpendicular retro-reflected laser beams at a wavelength $\lambda=1064.5$ nm. At the end of the ramp the lattice depth along the horizontal direction is $V_{x,y} = 30 E_R$, creating an ensemble of independent one-dimensional "tubes" with a transversal trap frequency $\omega_\perp = 2\pi \times 14.5$ kHz oriented along the vertical $z$-direction (Fig.~\ref{FIG1}(a)). Here, $E_R = h^2/(2 m \lambda^2)$ is the photon recoil energy with the mass $m$ of the Cs atom. During lattice loading the scattering length $a_s$ is set to $a_s = 173(5) a_0$ via a broad Feshbach resonance \cite{supmat}. In the deep lattice we then ramp $a_s$ within $50$ ms to the desired value in the range $10 a_0 \lesssim a_s \lesssim 900 a_0$ to prepare the tubes close to the adiabatic ground state. The ramp of $a_s$ is carefully adapted to avoid any excitation of breathing modes.

The gas in each tube is described by the Lieb-Liniger Hamiltonian \cite{Lieb63}
\begin{equation}
\hat{H} = -\frac{\hbar^2}{2 m} \sum \limits_i \partial^2 / \partial {z_i}^2 + g_{\rm{1D}} \sum \limits_{\langle i,j \rangle} \delta(z_i-z_j)\, ,
\label{EQ1}
\end{equation}
with $g_{\rm{1D}} = 2 \hbar \omega_\perp a_s \left( 1- 1.0326 \, a_s/a_\perp \right)^{-1}$ the coupling strength in 1D \cite{Haller09,Olshanii98,Haller10} and $a_\perp=\sqrt{\hbar/(m \omega_\perp)}$ the transverse harmonic oscillator length. The system is conveniently described in terms of the dimensionless interaction parameter $\gamma=m g_{\rm{1D}}/(\hbar^2 n_{\rm{1D}})$, where $n_{\rm{1D}}$ denotes the one-dimensional line density \cite{Cazalilla11}. The density sets the characteristic Fermi wave-vector $k_F=\pi n_{\rm{1D}}$ of the system. In our experimental setup we have to consider two sources of inhomogeneity. First, the tubes are harmonically confined along the longitudinal direction with a trap frequency $\omega_z= 2\pi \times 15.8(0.1)$ Hz. This gives rise to an inhomogeneous density distribution in each quantum wire. Second, the loading procedure leads to a distribution of the number of atoms across the ensemble of 1D systems \cite{supmat}. For comparing measurements with theoretical predictions both effects can be accounted for by averaging over homogeneous subsystems in a local density approximation (LDA) (see insets to Fig.~\ref{FIG1}(c) and (d)).

\begin{figure}
\begin{tabular}{cc}
\includegraphics[width=0.55\columnwidth]{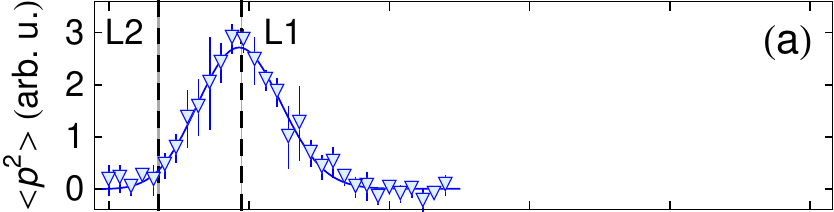} & \multirow{5}{*}[26.9pt]{\minitab[c]{\includegraphics[width=0.42\columnwidth]{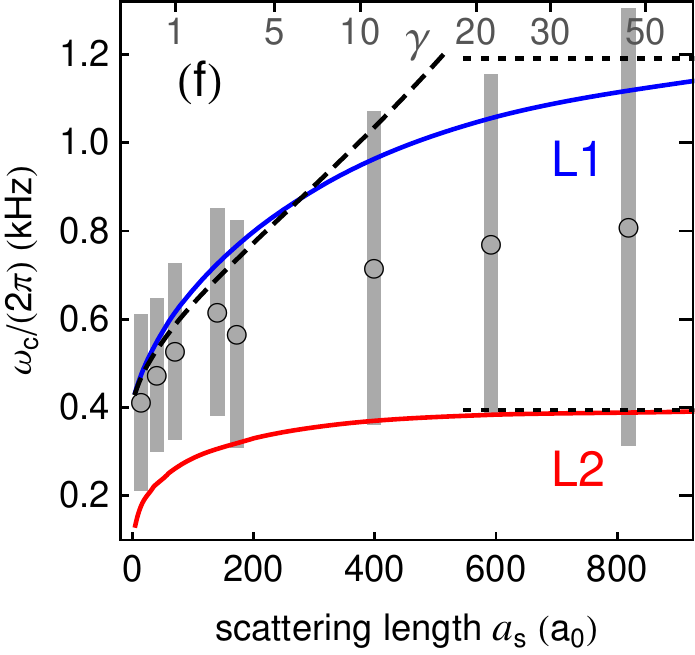} \\ [-0.2pt]\includegraphics[width=0.42\columnwidth]{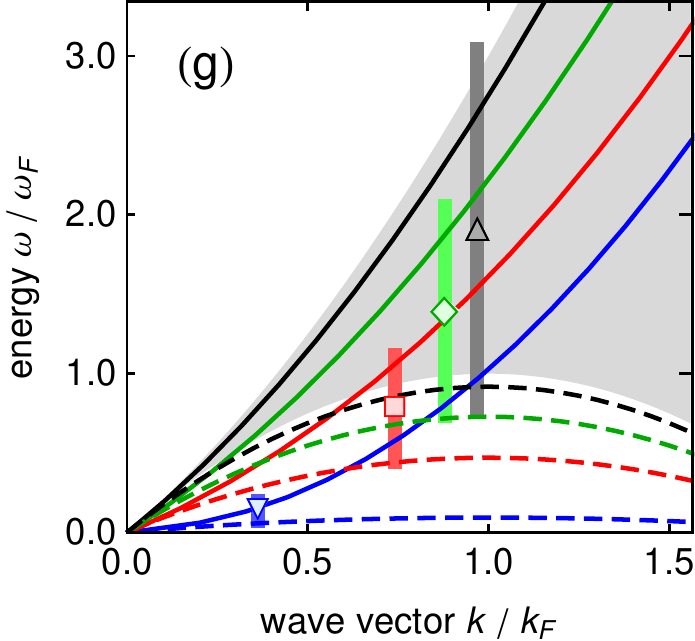}}} \\ [-1.2pt]
\includegraphics[width=0.55\columnwidth]{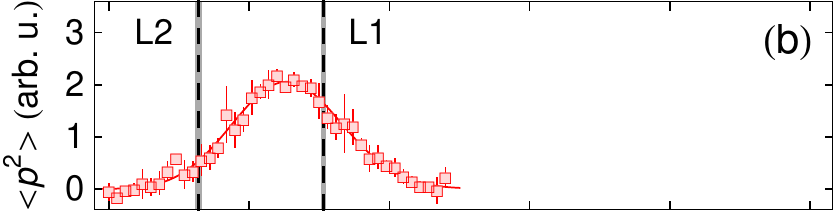} \\ [-1.2pt]
\includegraphics[width=0.55\columnwidth]{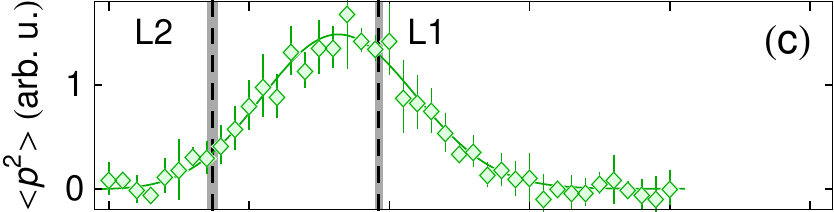} \\ [-1.2pt]
\includegraphics[width=0.55\columnwidth]{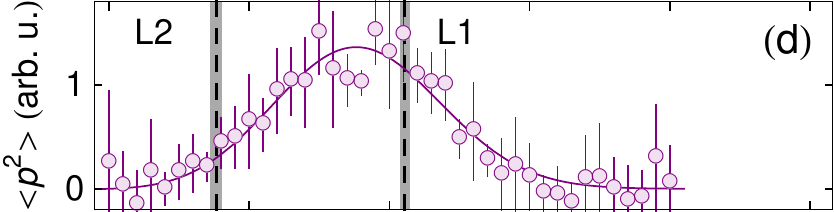} \\ [-1.2pt]
\includegraphics[width=0.55\columnwidth]{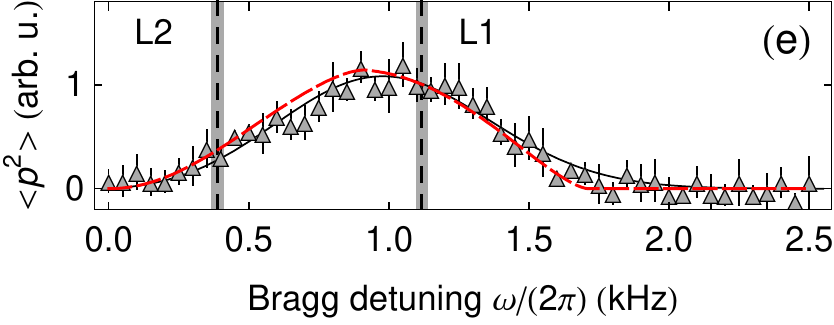}
\end{tabular}
\caption{\label{FIG2}(color online). Bragg-excitation spectra for different values of the 1D interaction strength $\gamma$. (a)-(e) Transferred energy $\sim \langle p^2 \rangle$ (normalized to unit area) as a function of the Bragg detuning $\omega$. The scattering length is set to 15$a_0$ (a), 173$a_0$ (b), 399$a_0$ (c),  592$a_0$ (d), and 819$a_0$ (e), giving an average $\gamma$ $(k_F / \mu \rm{m}^{-1})$ of 0.12(5) $(9(1))$, 3.3(1) $(4.4(2))$, 11.0(4) $(3.7(1))$, 21.7(7) $(3.5(1))$, and 45(1) $(3.3(1))$, respectively. Solid lines show fits to the data using a Gaussian multiplied by $\omega$. The vertical dashed lines indicate the position of the Lieb-I (L1) and Lieb-II (L2) mode calculated with the averaged values for $\gamma$. The dashed line in (e) shows the calculated response for ensemble-averaged trapped TG gases. (f) Central excitation energy $\omega_{\rm{c}}$ extracted from the Gaussian fit model as a function of $a_s$ (circles). Solid (dashed) lines denote the calculated position of the Lieb-I/II modes (Bogoliubov mode). The dotted lines indicate the energy of particle and hole excitations in the TG limit. (g) $\omega_{\rm{c}}$ extracted from the data shown in (a) (triangle), (b) (square), (c) (diamond), and (d) (inverted triangle) plotted in the dimensionless energy-momentum plane. Solid (dashed) lines denote the Lieb-I (Lieb-II) mode. The shaded area shows the continuum of excitations in the TG limit. Vertical lines in (f) and (g) give the fitted FWHM.}
\end{figure}

We probe the spectrum of elementary excitations via two-photon Bragg spectroscopy \cite{Stenger99}. In brief, the sample is illuminated for $5$~ms with a pair of phase coherent laser beams at a wavelength $\lambda_{\rm{B}} \approx 852$~nm and detuned by $\approx 200$ GHz from the Cs D2 line. The beams intersect at an angle $\phi$ at the position of the atoms and are aligned such that the wave-vector difference points along the direction of the tubes (Fig.~\ref{FIG1}(b)). Its magnitude $k = 4\pi / \lambda_{\rm{B}} \sin (\phi/2)$ sets the momentum transfer, while a small frequency detuning $\omega$ between the laser beams defines the energy transfer to the system. In linear response, the energy absorbed from the Bragg lasers for a fixed pulse area $\Delta E (k,\omega)$ directly relates to the dynamical structure factor $S(k,\omega) = \int dx \int dt e^{i\omega t - i k x} \langle \rho(x, t) \rho(0, 0) \rangle$ at finite temperature $T$ via $\Delta E (k,\omega) \propto \hbar \omega (1-e^{- \hbar \omega / (k_B T)}) S(k,\omega)$ with Boltzman's constant $k_B$ \cite{Brunello01}.

\begin{figure*}
\includegraphics[width=0.48\columnwidth]{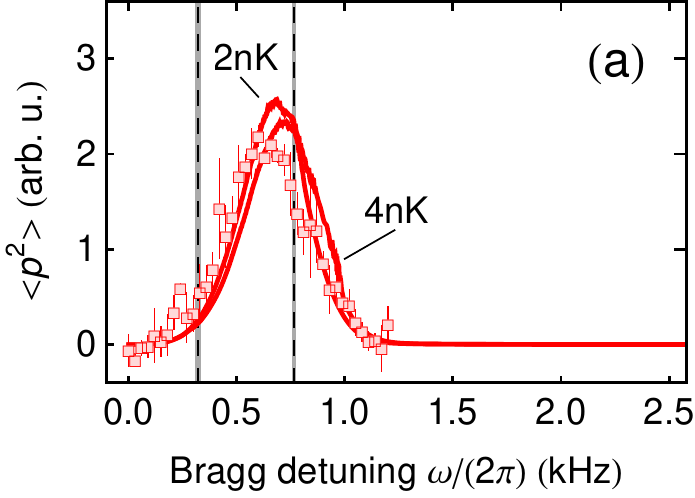}
\hspace{1mm}
\includegraphics[width=0.48\columnwidth]{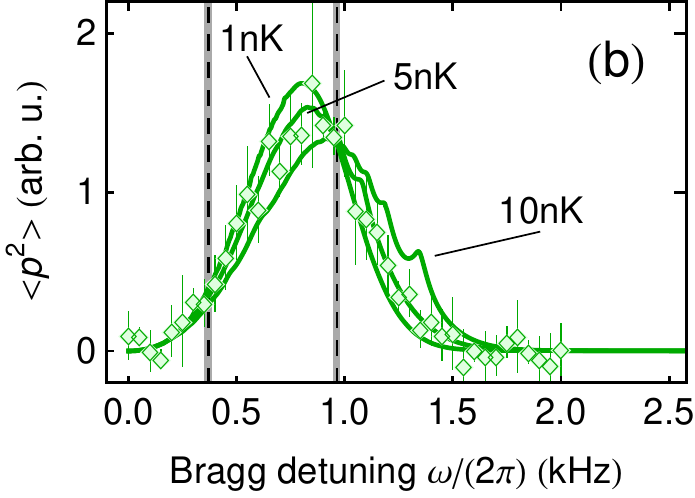}
\hspace{1mm}
\includegraphics[width=0.48\columnwidth]{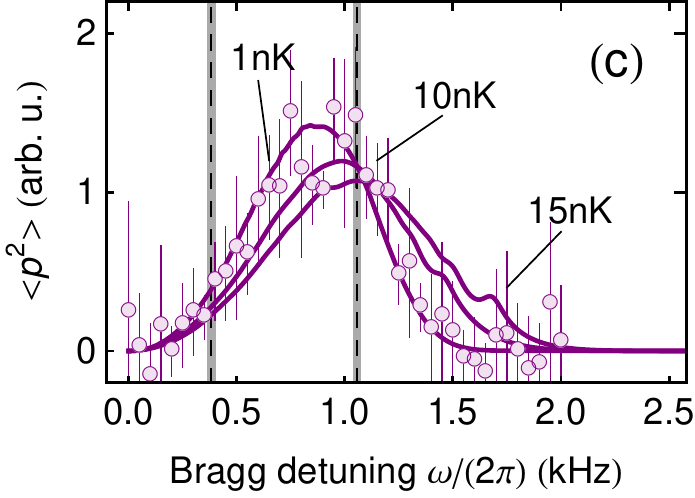}
\hspace{1mm}
\includegraphics[width=0.48\columnwidth]{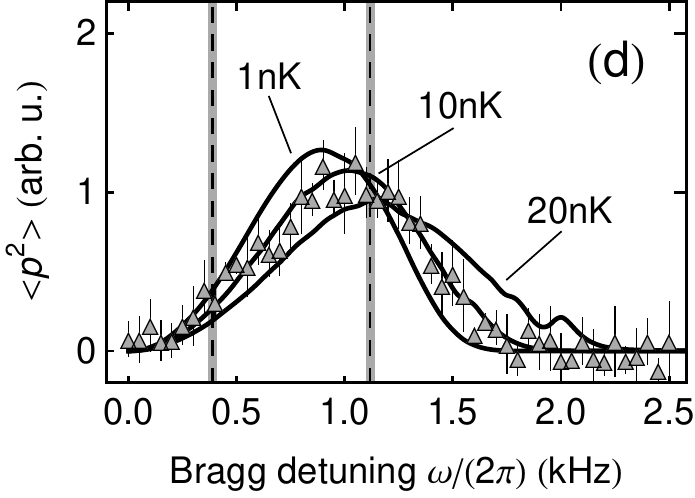}\\
\vspace{1mm}
\includegraphics[width=0.48\columnwidth]{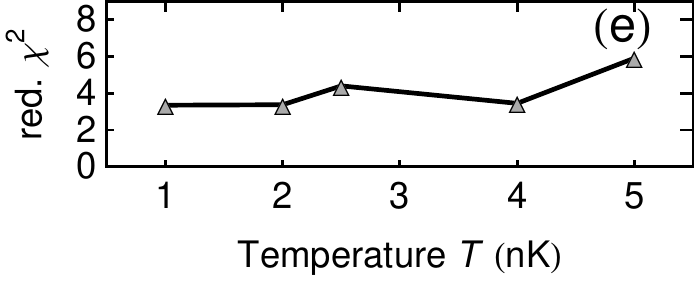}
\hspace{1mm}
\includegraphics[width=0.48\columnwidth]{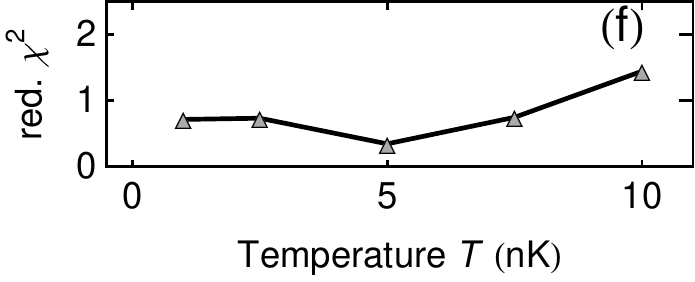}
\hspace{1mm}
\includegraphics[width=0.48\columnwidth]{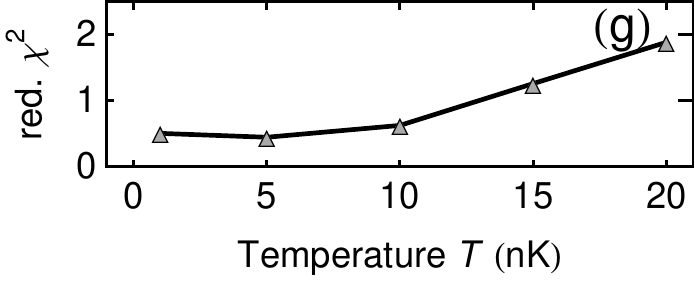}
\hspace{1mm}
\includegraphics[width=0.48\columnwidth]{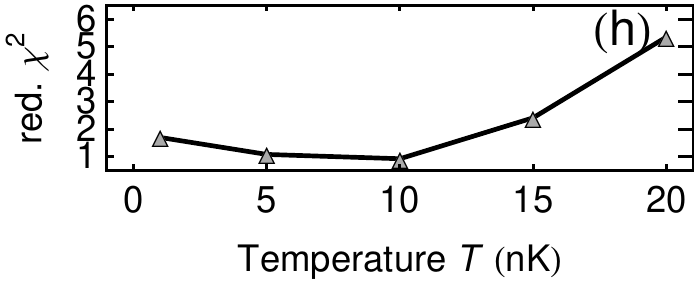}
\caption{\label{FIG3}(color online). Comparison of the Bragg-excitation spectra with theoretical predictions at finite temperature. Symbols depict the data presented in Fig.~\ref{FIG2} for $\gamma=$3.3(1) (a), 11.0(4) (b),  21.7(7) (c), and 45(1) (d). The vertical dashed lines indicate the position of the Lieb-I and Lieb-II modes calculated with the averaged values for $\gamma$. The solid lines show the ensemble averaged response derived from the finite temperature dynamical structure factor. (e)-(h) Reduced $\chi^2$ analysis of the corresponding experimental data in the top row with theoretical predictions for different temperatures.}
\end{figure*}

In our experiment, we probe the ensemble of 1D tubes at a fixed $k=3.24(3) \, \mu \rm{m}^{-1}$, which is comparable to typical mean values for $k_F$ averaged over the sample \cite{calib}. The absorbed energy as a function of $\omega$ is measured in momentum space. For this, we switch off the lattice potential quickly (within $300 \, \mu \rm{s}$) and allow for 50 ms time-of-flight at a small positive scattering length of $\approx 15 a_0$. From the integrated line density along the $z$-direction of the tubes we extract $\langle p^2 \rangle$ and plot it as a function of $\omega$. The result for five different values of $\gamma$ is depicted in Fig.~\ref{FIG2}(a)-(e).

The datasets cover the regime from weak to strong interactions $0.1 \lesssim \gamma \lesssim 50$ probed at  $0.3 \lesssim k/k_F \lesssim 1$. The variation in $k/k_F$ ensues from the change of the density distribution in the tubes with increasing $a_s$, evolving from a Thomas-Fermi profile at weak interactions towards the TG profile at strong interactions. The values for $\gamma$ and $k_F$ given in Fig.~\ref{FIG2} denote the average over the ensemble of 1D systems using the mean $n_{\rm{1D}}$ in each wire calculated for $T=0$ from the solution of the Lieb-Liniger integral equations in LDA \cite{supmat}. Error bars reflect mainly a $\pm10\%$ uncertainty in the total atom number. A clear interaction-induced broadening and shift of the spectra with increasing $\gamma$ is observed in accordance with the calculated position of the Lieb-I and Lieb-II modes (vertical dashed lines) \cite{Lieb63b}. We compare the dataset in the strongly interacting regime (Fig.~\ref{FIG2}(e)) to the calculated response for an ensemble of trapped TG gases at zero temperature averaged over the ensemble of tubes (dashed line) \cite{Golovach09,supmat}. The agreement with the data underlines the contribution of Lieb's hole-like excitation to the dynamical response.

Prior to a detailed discussion on the exact lineshape for finite $\gamma$ and finite $T$, we attempt a simplified zero-temperature analysis of our spectroscopy signal. A function $\propto \omega \, \cal{G}(\omega_{\rm{c}})$ is fit to the data (solid lines), where the DSF averaged over the ensemble of tubes is approximated by a simple Gaussian function $\cal{G}$ centered at $\omega_{\rm{c}}$. The extracted $\omega_{\rm{c}}$ as a function of $a_s$ is shown in Fig.~\ref{FIG2}(f) (circles). The vertical lines denote the fitted full width at half maximum (FWHM). We find the spectral weight of the data within the Lieb-I and Lieb-II modes (solid lines) calculated using the ensemble averaged values for $\gamma$. Bogoliubov's quasi particle energy alone (dashed line) does not explain the observed $\omega_{\rm{c}}$ with increasing $\gamma \gtrsim 3$. We summarize the measured spectral position and width for four selected values of $\gamma$ in the dimensionless energy-momentum plane in Fig.~\ref{FIG2}(g), with $\omega_F=\epsilon_F/\hbar=\hbar k_F^2/(2m)$. Comparing to the dispersion relation of Lieb's particle and hole-like excitation, we observe a clear signature for the contribution of both branches with increasing interactions, finally approaching the limit of the excitation spectrum of the fully fermionized Bose gas (shaded area).

We now turn to a detailed analysis of the spectral lineshape as a function of interaction strength and temperature. The effect of temperature on the measurement of the DSF arises from two distinct contributions. First, the DSF itself is temperature dependent. This becomes most evident in the TG limit of fermionized bosons. For $k_B T \ll \epsilon_F$ the effective Fermi sea in quasi-momentum space has a sharp edge giving rise to a homogeneous continuum of excitations lying between Lieb's hole- and particle-like modes. When $k_B T \sim \epsilon_F$ the Fermi edge washes out, resulting in a smoothening of the DSF with its spectral weight being shifted to higher energies \cite{Panfil14}. Second, finite temperature affects the density distribution in our 1D systems leading to a decreasing mean $n_{\rm{1D}}$ with increasing $T$ for fixed $a_s$. This changes the average $k/k_F$ at which the tubes are probed. 

For our theoretical analysis we take both effects into account. First, we calculate the density distribution in each of the tubes at a temperature $T$ by numerically solving the Yang-Yang integral equations for the 1D Bose gas \cite{Yang69}. The DSF is evaluated at finite $T$ via the ABACUS algorithm, a Bethe Ansatz-based method to compute correlation functions of integrable models~\cite{Caux09}. The effect of the trapping potential is incorporated by making a LDA for each tube. The response of the array of tubes is finally calculated by weighting the contribution of each subsystem by the number of atoms \cite{supmat}.

\begin{figure}[!ht]
\begin{tabular}{cc}
\includegraphics[width=0.48\columnwidth]{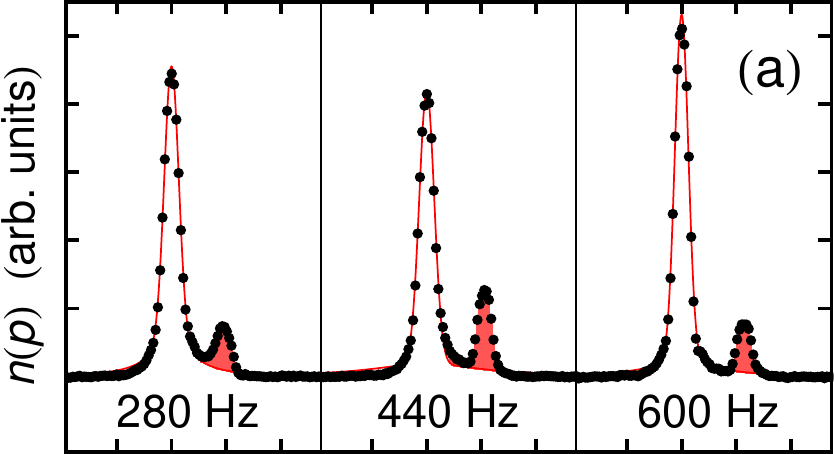} & \multirow{3}{*}[57pt]{\minitab[c]{\includegraphics[width=0.48\columnwidth]{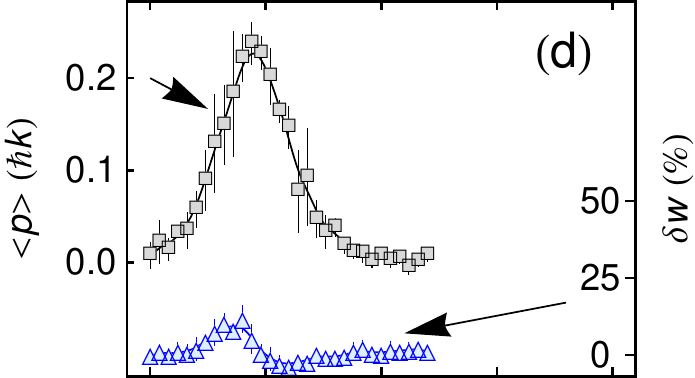} \\ [-2.5pt]
\includegraphics[width=0.48\columnwidth]{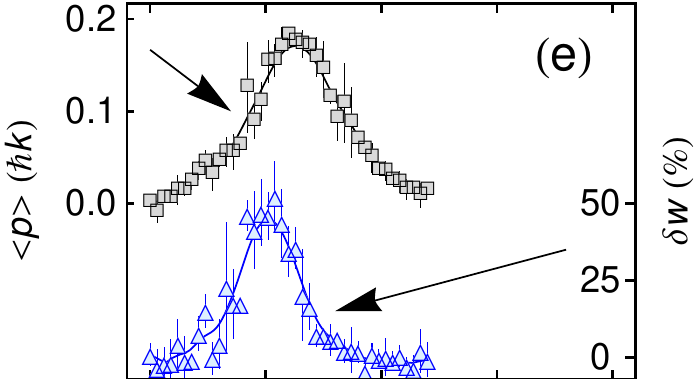} \\ [-2.5pt]
\includegraphics[width=0.48\columnwidth]{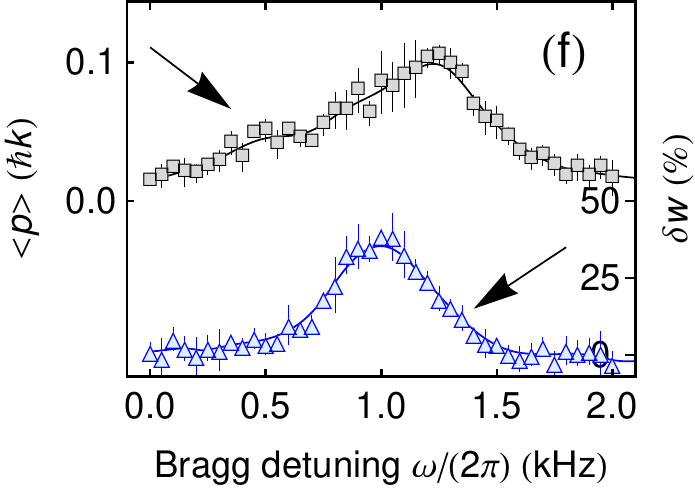}}} \\ [-2.5pt]
\includegraphics[width=0.48\columnwidth]{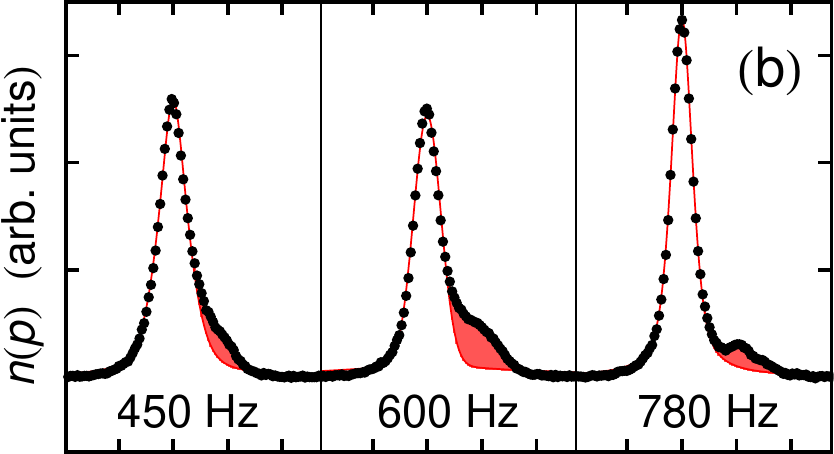} \\ [-2.5pt]
\includegraphics[width=0.48\columnwidth]{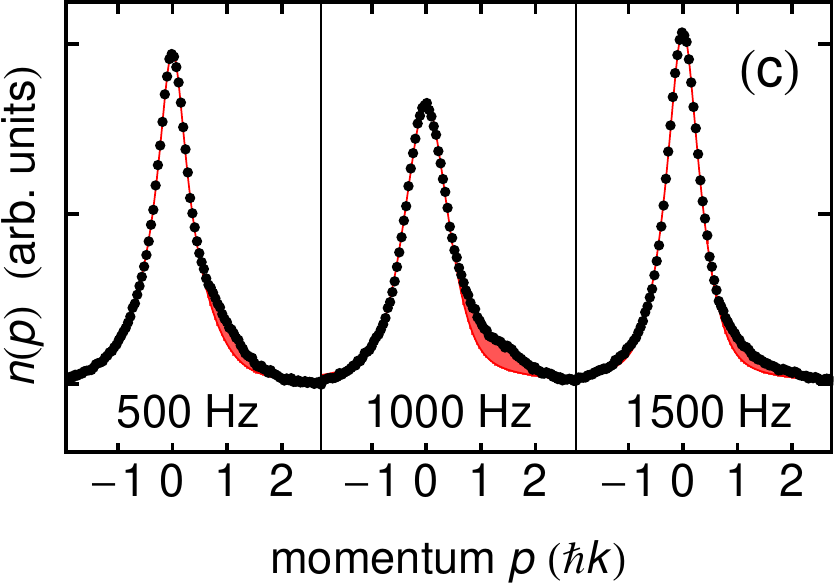}
\end{tabular}
\caption{\label{FIG4}(color online). Momentum distribution $n(p)$ for resonant and off-resonant Bragg excitation measured at weak and strong interactions. (a)-(c) Integrated line density after 50 ms time-of-flight when exciting the 1D systems at $\gamma=$ 0.12(5) (a), 3.3(1) (b), 45(1) (c) for red-detuned (left), resonant (middle) and blue-detuned (right) excitation with respect to the Bragg resonance. Datapoints show the average of four measurements. The frequencies given denote the Bragg detuning $\omega/(2\pi)$. The red line shows a fit to the central part of the line density around $p=0$ using a linear combination of a Lorentzian and Gaussian profile to extract the width $w$. The red shaded area denotes the fraction of atoms asymmetrically scattered to higher momentum states. (d)-(f) Imparted mean momentum $\langle p \rangle$ extracted from $n(p)$ (squares) and relative change in the fitted width $\delta w$ of the central peak around $p=0$ (triangles) as a function of the detuning $\omega/(2\pi)$ for $\gamma=$ 0.12(5) (d), 3.3(1) (e), and 45(1) (f). Solid lines are interpolating functions to guide the eye.}
\end{figure}

The result of our theoretical analysis for four different values of $\gamma$ is presented in Fig.~\ref{FIG3} and compared to the corresponding experimental data (taken from Fig.~\ref{FIG2} (b-e)). Although finite temperature leads to small shifts and broadening of the excitation spectrum, the most relevant contribution to the spectral shape stems from the broadening of the dynamical structure factor with increasing interactions. The analysis underlines the contribution of hole-like excitations to the overall response when entering deep into the strongly correlated regime. Further, a reduced $\chi^2$ analysis of our data with the computed spectra serves as a thermometry tool in the tubes and points to gas temperatures in the range of 5 to 10 nK. A moderate increase in temperature is seen for increasing values of $\gamma$ \cite{supmat}.

So far, we have characterized the excitations in the gas by measuring $\langle p^2\rangle$. Now, we analyze the full momentum distribution $n(p)$ of the excited 1D Bose gases. In Fig.~\ref{FIG4}(a)-(c) we plot the atomic line density after time-of-flight, integrated transversally to the direction of the tubes, which reflects the in-trap momentum distribution of the atomic ensemble. The measurements are taken at three different values of $\gamma$, ranging from the weakly to the strongly interacting regime, and at Bragg excitation frequencies $\omega$ slightly below (left column), just at (central column), and slightly above (right column) the peak of the resonance.

First, we recognize a dramatic qualitative change in $n(p)$ with increasing $\gamma$ for all detunings presented. In the weakly interacting regime (Fig.~\ref{FIG4}(a)), we observe a clear particle-like excitation located at $p=\hbar k$ as expected from the non-interacting limit. Yet, with increasing interactions this feature smears out and evolves towards an overall broadening of $n(p)$, indicative of a strong collective response of the system to the Bragg pulse. This observation demonstrates one of the key features of 1D systems: any excitation to the system is necessarily collective, and therefore leads to energy-broadened response functions, in contrast to sharp coherent single-particle modes. This broadening however only becomes clearly visible for strong enough interactions, where the hole-like modes become dynamically relevant.

In a further measurement, we attempt to quantify the response in momentum space in more detail. Note that our previous measurement of $\langle p^2\rangle$ captures both a broadening of the momentum distribution as well as an increase in the mean momentum $\langle p\rangle$. In order to separate both contributions spectroscopically, we plot the relative change $\delta w$ of the width $w$ of the central part of $n(p)$ around $p=0$ and $\langle p\rangle$ as a function of $\omega$ for different values of $\gamma$ in Fig.~\ref{FIG4}(d)-(f). The data indicate how collective excitations in the gas, expressed via energy deposition in $w$ rather than in $\langle p\rangle$, become dominant with increasing $\gamma$. Interestingly, the two curves peak at different values for $\omega$. This observation is confirmed by the momentum-space character of elementary excitations in the interacting 1D systems, which changes from a collective broadening for the Lieb-II mode to a particle-like feature for the Lieb-I mode with increasing $\omega$ \cite{supmat}. 

In conclusion, we have measured the excitation spectrum of a strongly correlated 1D system for a wide range of interaction parameters. Comparison with integrability-based calculations at finite temperature allows for a direct observation of the contribution of the collective Lieb-II mode to the DSF. Our results demonstrate the successful application of an integrable model to analyze dynamics of the 1D Bose gas. Furthermore, the collective nature of elementary excitations in 1D with increasing interaction strength has been demonstrated through an analysis of the momentum distribution. Our results pose questions on the time evolution and potential equilibration of these collective excitations. This could be seen as an alternative quantum cradle setting \cite{Kinoshita06} in which, instead of colliding two highly energetic clouds of atoms, relatively low energy excitations propagate through the system, whose individual features can then be more easily studied.

We are indebted to R. Grimm for generous support, and thank M. Buchhold and S. Diehl for fruitful discussions. We gratefully acknowledge funding by the European Research Council (ERC) under Project No. 278417 and under the Starting Grant No. 279391 EDEQS, by the Austrian Science Foundation (FWF) under Project No. P1789-N20, and from the FOM and NWO foundations of the Netherlands.

\bibliographystyle{apsrev}

\clearpage

\section{Supplementary Material: Probing the Excitations of a Lieb-Liniger Gas from Weak to Strong Coupling}
 
\subsection{Lattice depth calibration and error bars}

The lattice depth $V_{x,y}$ is calibrated via Kapitza-Dirac diffraction. The statistical error for $V_{x,y}$ is 1\%, though the systematic error can reach up to 5\%.

The scattering length $a_s$ is calculated via its dependence on the magnetic field \cite{Mark11MAT} with an estimated uncertainty of $\pm 5 a_0$ arising from systematics in the magnetic field calibration and conversion accuracy. Additionally, the magnetic field gradient for sample levitation \cite{Weber2002MAT,Kraemer2004MAT} leads to a variation of less than $\pm 3 a_0$ across the sample. 
 
\subsection{Atom number distribution across the 1D tubes}
 
\begin{figure}
\includegraphics[width=0.48\columnwidth]{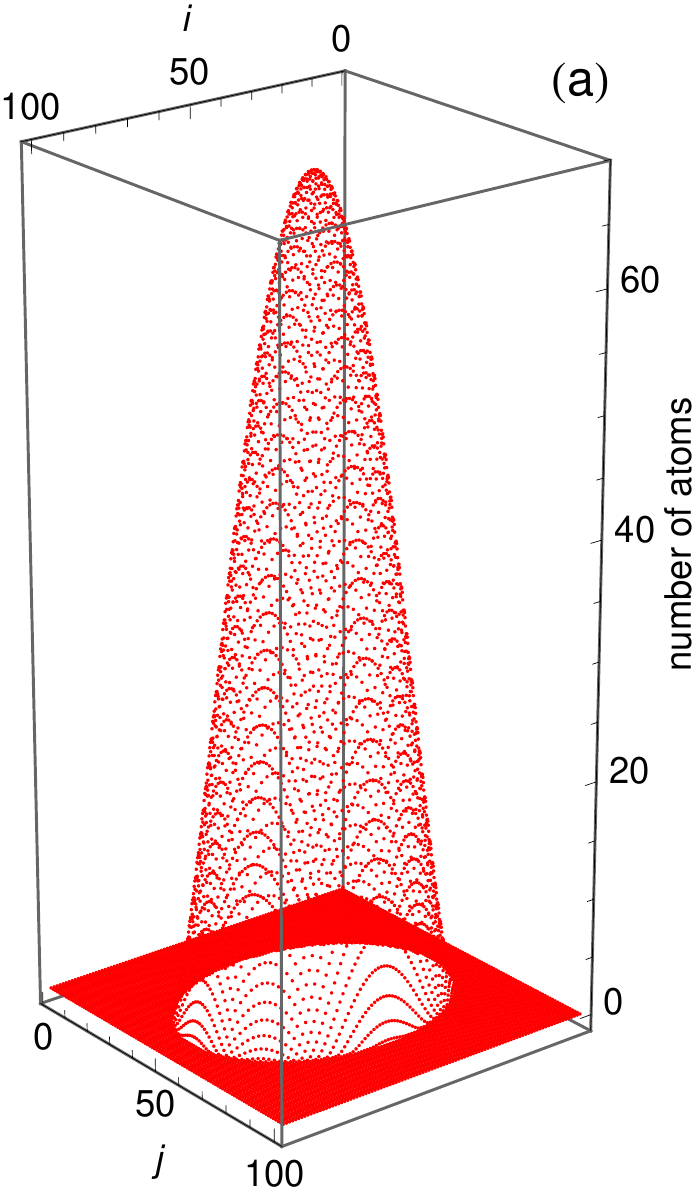}
\hspace{2mm}
\includegraphics[width=0.48\columnwidth]{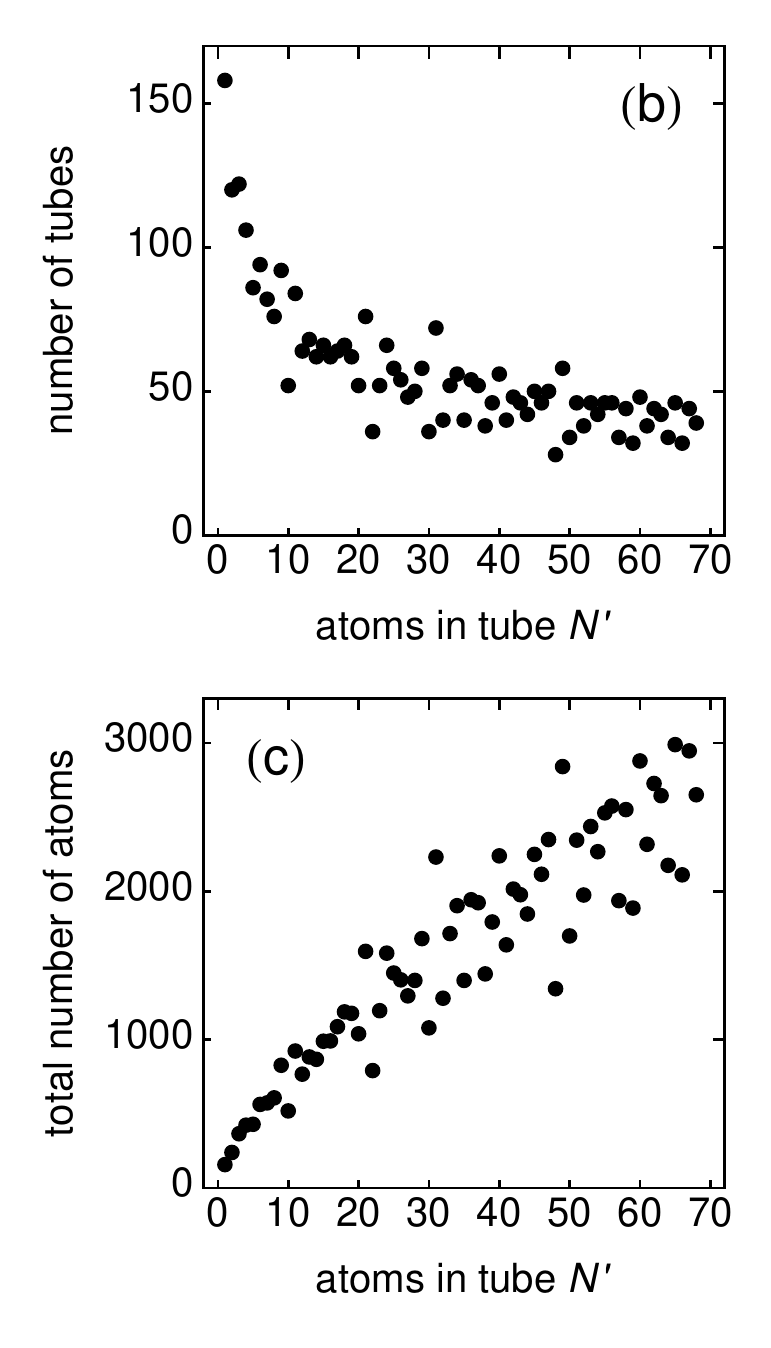}
\caption{\label{SuppFIG1} (a) Atom number distribution over the 1D tubes for a total atom number $N=1.1 \times 10^5$ in the BEC and a 3D scattering length during loading $a_{\rm{3D}}=173a_0$. (b) Number of tubes that are filled with $N_{i,j} = N'$ atoms. (c) Total number of atoms in tubes filled with $N_{i,j} = N'$ atoms.}
\end{figure}
 
In order to calculate a mean value of the interaction parameter $\gamma$ and the Fermi wave vector $k_F$ we need to know the atom number distribution across the ensemble of tubes, as shown in Fig.~\ref{SuppFIG1}. We follow the calculation presented in \cite{Haller11MAT}.

The BEC is adiabatically loaded from a crossed dipole trap into the optical lattice. When the lattice is fully ramped up to its final depth $V_{x,y} = 30 E_R$, the laser beams forming the array of tubes give rise to an additional background harmonic confinement. The total background harmonic confinement of lattice and dipole trap laser beams is measured to $\omega_x = 2 \pi \times 15.4(0.1) \, \rm{Hz}$, $\omega_y = 2 \pi \times 20.1(0.1) \, \rm{Hz}$, and $\omega_z = 2 \pi \times 15.8(0.1) \, \rm{Hz}$. We deduce the atom number $N_{i,j}$ for the tube $(i,j)$ from the global chemical potential $\mu$. Assuming that interactions are sufficiently small during the loading process so that all tubes are in the 1D Thomas-Fermi (TF) regime, the local chemical potential in each tube reads
$$
\mu_{i,j} = \mu - \frac{1}{2} m (\lambda/2)^2 \left( \omega_x^2 i^2 + \omega_y^2 j^2\right) \, ,
$$
with $m$ the mass of the Cs atom and $\lambda=1064.5 \, \rm{nm}$ the wavelength of the lattice light. From $\mu_{i,j}$ the atom number in each tube can be derived via
$$
\mu_{i,j} = \left (\frac{3 N_{i,j}}{4 \sqrt{2}} g_{\rm{1D}} \omega_z \sqrt{m} \right)^{2/3} \, ,
$$
with $g_{\rm{1D}}$ the 1D coupling strength
$$
g_{\rm{1D}} = 2 \hbar \omega_\perp a_s \left( 1- 1.0326 \frac{a_s}{a_\perp} \right)^{-1} \, .
$$
Here, $a_\perp = \sqrt{\hbar / (m \omega_\perp)}$ is the radial oscillator length. The global chemical potential is calculated iteratively from the condition $N = \sum_ {i,j} N_{i,j} ( \mu )$. \\ 

\begin{figure}
\includegraphics[width=0.48\columnwidth]{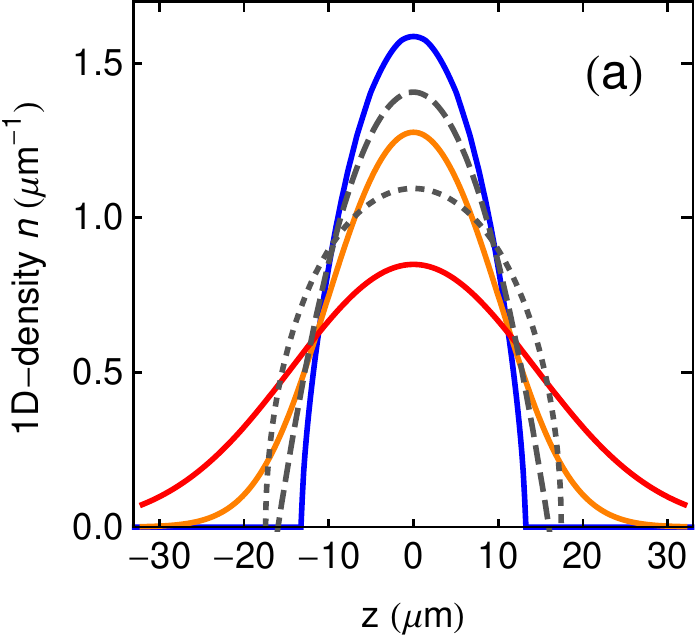}
\hspace{2mm}
\includegraphics[width=0.48\columnwidth]{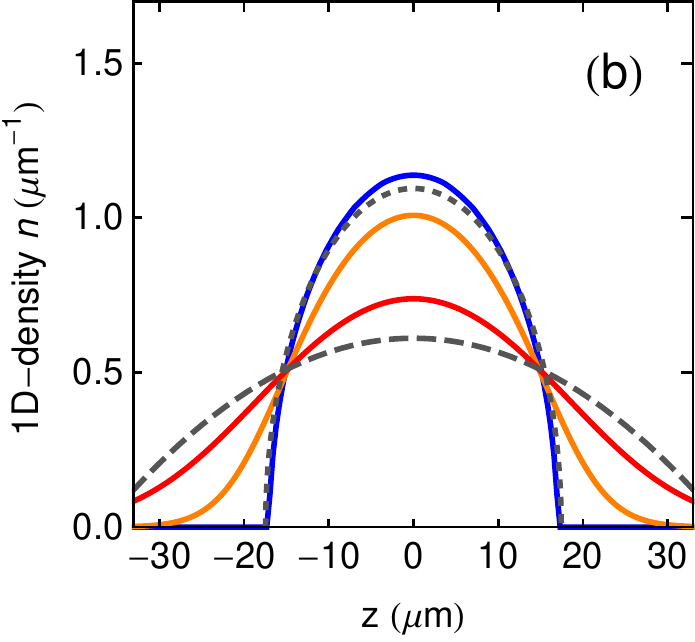}\\
\caption{\label{SuppFIG2} 1D line-density distribution of a tube with $N=30$ atoms at $a_s=150 a_0$ (a) and $a_s=800 a_0$ (b). The blue line shows the result from the numerical solution of the Lieb-Liniger system. The dashed (dotted) gray lines depict the analytic results for the profile in the Thomas-Fermi (Tonks-Girardeau) limit \cite{Dunjko01MAT}. Here, $\omega_\perp=2\pi \times 15$ kHz and $\omega_z=15$ Hz. Finite temperature profiles using the Yang-Yang thermodynamic equations are shown for $T=10$nK (orange line) and $T=30$nK (red line).}
\end{figure}

\subsection{Density profile in the tubes}

For $T=0$ we model the 1D density distribution $n(z)$ in each tube individually by numerically solving the Lieb-Liniger system and making a local density approximation \cite{Lieb63MAT,Dunjko01MAT}. For $T>0$ we solve the Yang-Yang thermodynamic equations of the 1D Bose gas making a local density approximation to calculate $n(z,T)$ \cite{Yang69MAT}. Representative examples for two different values of $a_s$ at zero and finite temperature are shown in Fig.~\ref{SuppFIG2}.

\subsection{Mean $\gamma$ and $k_F$}
 
\begin{figure}
\includegraphics[width=0.48\columnwidth]{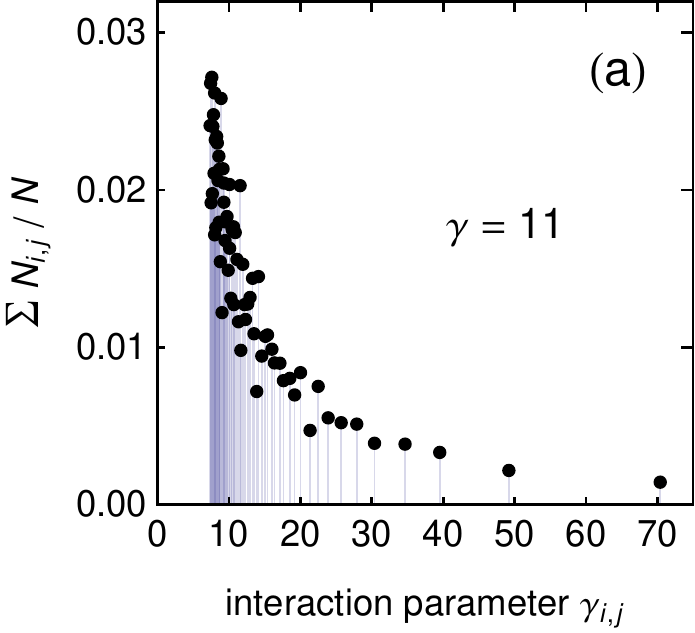}
\hspace{2mm}
\includegraphics[width=0.48\columnwidth]{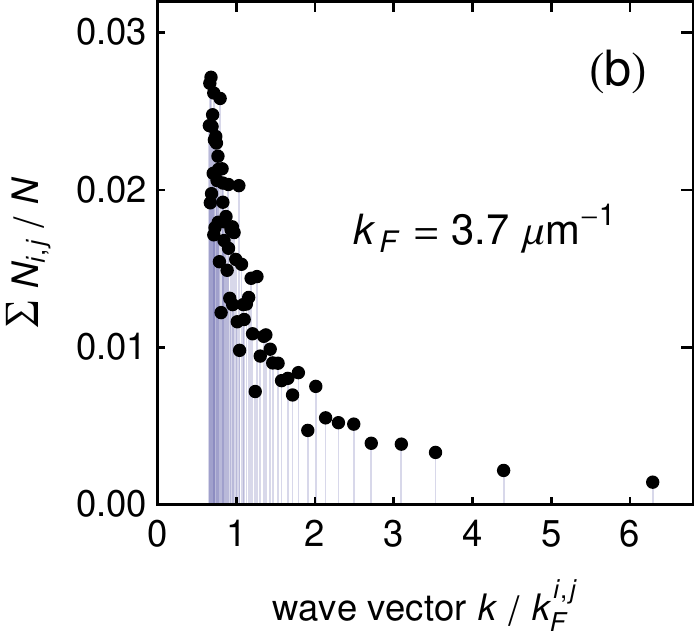}\\
\vspace{2mm}
\includegraphics[width=0.48\columnwidth]{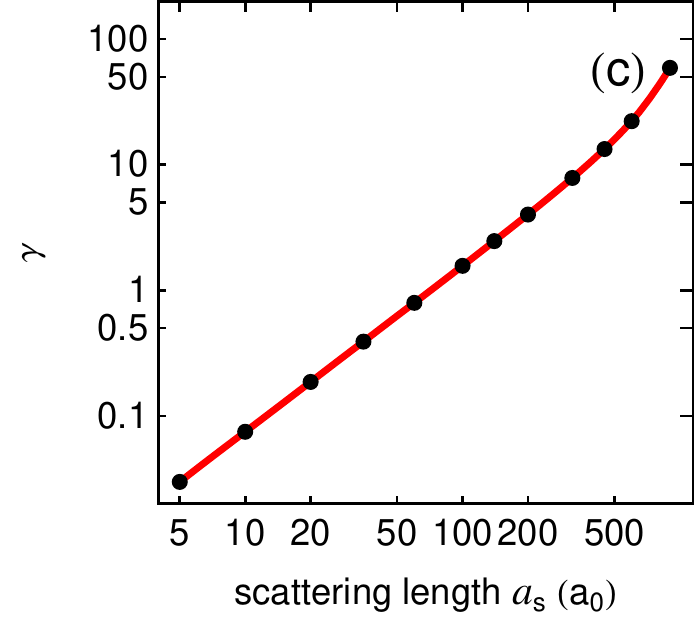}
\hspace{2mm}
\includegraphics[width=0.48\columnwidth]{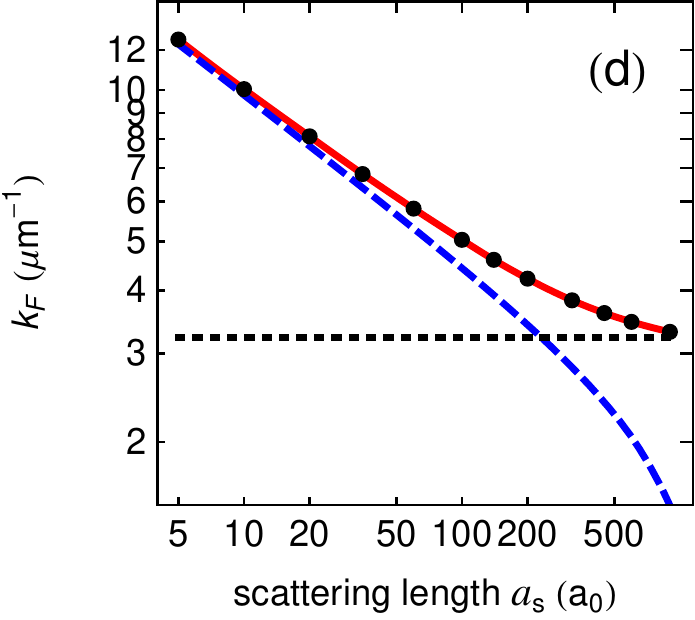}
\caption{\label{SuppFIG3}(top row) Relative atom number in tubes with local interaction parameter $\gamma_{i,j}$ (a) and a local value for $k/k_F^{i,j}$ (b). As an example, the distributions are calculated for $a_s = 399 a_0$ and $N=1.1 \times 10^5$ resulting in a global  $\gamma \, (k_F)$ of 11 $(3.7 \mu \rm{m}^{-1})$. (bottom row) Global values for $\gamma$ (c) and $k_F$ (d) as a function of the scattering length $a_s$. The points show the calculated values with $n_{\rm{1D}}$ computed numerically within the Lieb-Liniger model using a local density approximation. The dashed line in (d) is calculated with the density in the Thomas-Fermi limit while the dotted line shows the result using the density for a Tonks-Girardeau gas. The solid lines are interpolation functions to guide the eye. In (a)-(d) all quantities have been calculated using the mean value for $n_{\rm{1D}}$ in each tube.}
\end{figure}

From the atom number distribution, we calculate the density profile for each tube individually as described above. The mean 1D density in each tube $n_{i,j}^{\rm{1D}}$ then delivers local and mean values for $\gamma$ and $k_F$
$$
\gamma_{i,j} = \frac{m g_{\rm{1D}}}{\hbar^2 n_{i,j}^{\rm{1D}}} \, , \qquad \gamma = \frac{1}{N} \sum\limits_{i,j} N_{i,j} \gamma_{i,j}
$$
and
$$
k_F^{i,j} = \pi n_{i,j}^{\rm{1D}} \, , \qquad k_F = \frac{1}{N} \sum\limits_{i,j} N_{i,j} k_F^{i,j} \, .
$$
As an example, the distribution of atoms over tubes with local $\gamma_{i,j}$ and $k/k_F^{i,j}$ is depicted in Fig.~\ref{SuppFIG3}(a) and (b) for a specific value of $a_s$. In Fig.~\ref{SuppFIG3}(c) and (d) we show the mean value of $\gamma$ and $k_F$ as a function of $a_s$.

\subsection{Sampling of the dynamical structure factor over the array of tubes and comparison with the experimental data}

In linear response, the energy $\Delta E (k,\omega,T)$ dumped into a single 1D gas after the Bragg pulse relates to the dynamical structure factor $S(k,\omega,T)$ via \cite{Pitaevskii03MAT}
$$
\Delta E (k,\omega,T) \propto \hbar \omega (1-e^{- \hbar \omega / (k_B T)}) S(k,\omega,T) \, .
$$
Provided that the system is probed at sufficiently large momentum $k$ the DSF of the trapped system can be computed in a local density approximation
$$
S(k,\omega,T) = \frac{1}{2L} \int \limits_{-L}^{L} S_{\rm{hom}}(k,\omega,T ; n(z)) dz \, ,
$$
with $L$ the system length and $S_{\rm{hom}}$ the DSF calculated for a homogeneous system with uniform density. In practice, we find that dividing the profile of each tube in $\approx 10$ homogeneous subsystems approximates the DSF of the trapped gas sufficiently well. The DSF in each tube with $N_{i,j}$ atoms obeys the $f$-sum rule \cite{Pitaevskii03MAT},
$
\int \limits_{-\infty}^{+\infty} \omega S(k, \omega, T) \, d \omega \propto N_{i,j} \, . 
$
In combination with detailed balance $S(k,\omega)=e^{\hbar \omega / (k_B T)} S(k,-\omega)$ we find
$$
\int \limits_{0}^{+\infty} \omega S(k, \omega, T) (1- e^{- \hbar \omega / (k_B T)} ) \, d \omega \propto N_{i,j} \, . 
$$
This allows us to calculate the dynamical response of the entire ensemble by first normalizing the DSF of each individual tube $(i,j)$ by its $f$-sum, respectively, and then weighting its contribution to the total signal by the number of atoms $N_{i,j}$. For a direct comparison with the experimental signal, we finally normalize the ensemble averaged response to unit area. \\

In the experiment, we extract $\langle p^2 \rangle$ as a function of $\omega$ from the momentum space distribution obtained after time-of-flight. For a direct comparison, each spectrum is normalized to unit area as the overall signal depends on the intensity of the Bragg lasers, which we slightly increase for data taken at stronger interactions. A changing offset due to an overall broadening of the momentum distribution with increasing $\gamma$ is subtracted from the data. This offset does not affect the shape of the excitation spectrum and stems from the broadening of the unperturbed momentum distribution with increasing $\gamma$.

\begin{figure}
\includegraphics[width=0.48\columnwidth]{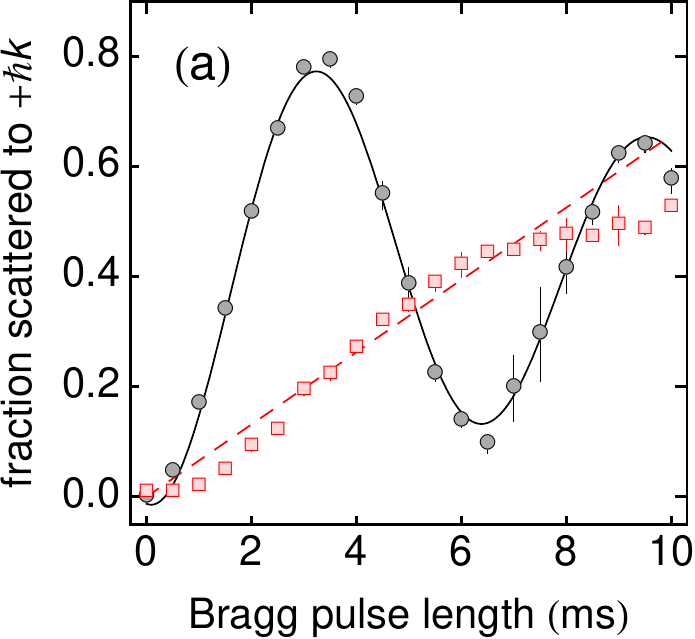}\\
\vspace{2mm}
\includegraphics[width=0.48\columnwidth]{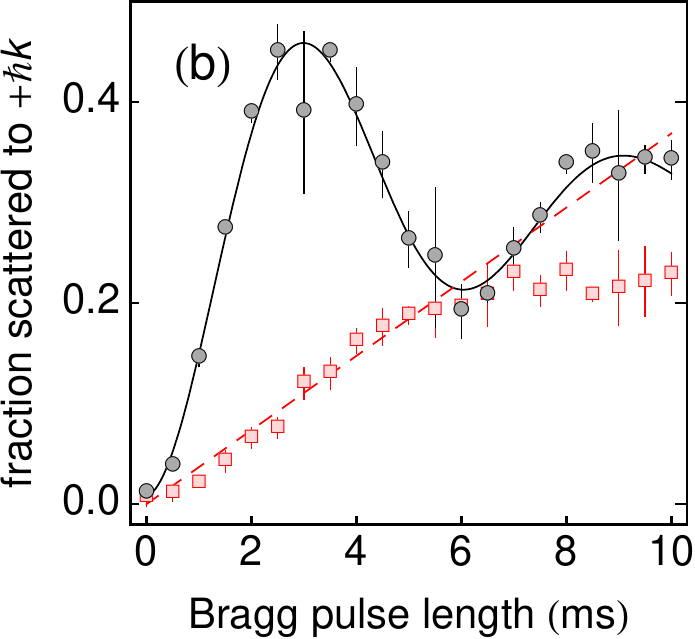}
\hspace{2mm}
\includegraphics[width=0.48\columnwidth]{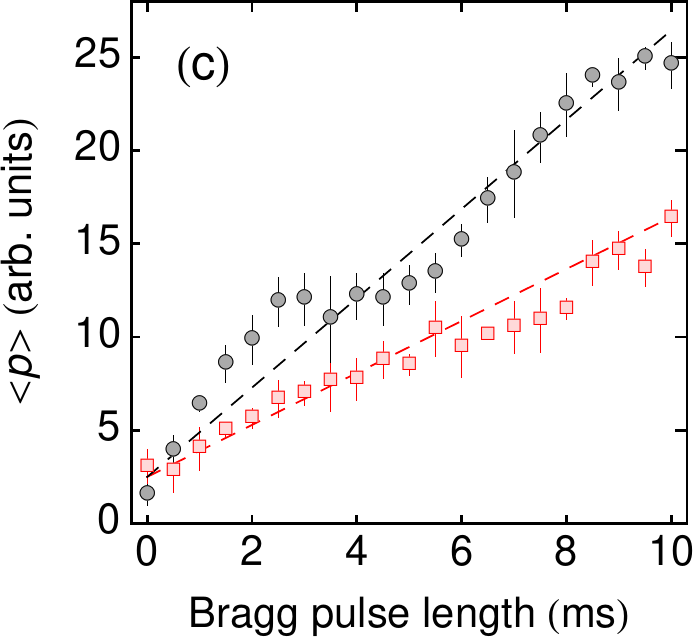}\\
\vspace{2mm}
\includegraphics[width=0.48\columnwidth]{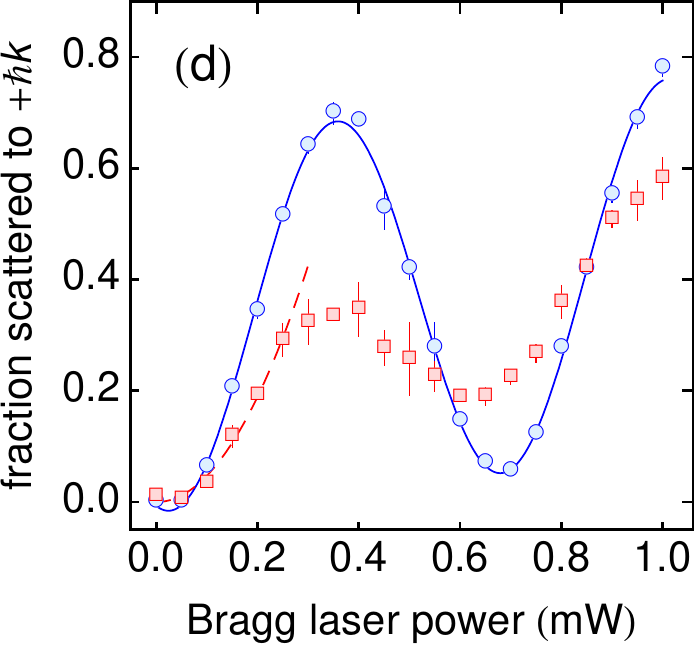}
\hspace{2mm}
\includegraphics[width=0.48\columnwidth]{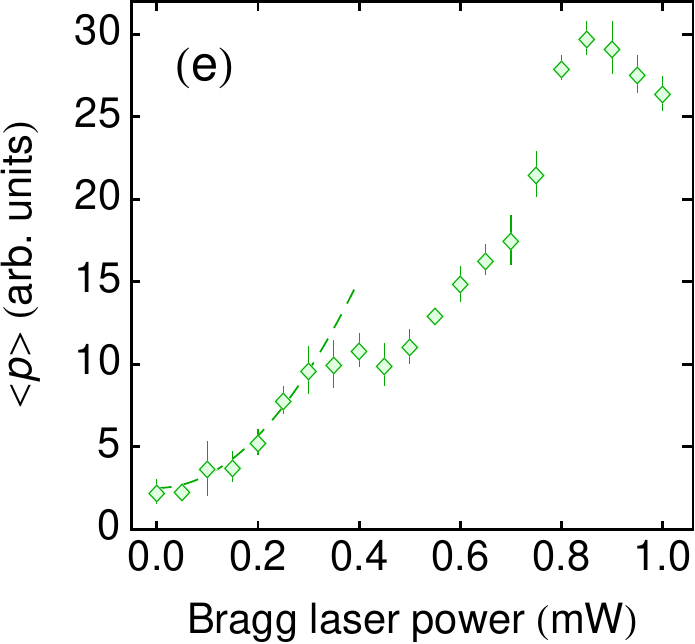}
\caption{\label{SuppFIG4}(a)-(c) Bragg excitation as a function of the laser pulse length measured in a weakly interacting 3D BEC (a), in 1D at $a_s = 15 a_0$ (b), and in 1D at $a_s = 819 a_0$ (c). The laser power in both Bragg beams is set to $0.5$mW (circles) and $0.2$mW (squares) in (a) and (b), and $0.5$mW (circles) and $0.3$mW (squares) in (c). (d),(e) Bragg excitation as a function of the laser power in both Bragg beams measured in a weakly interacting BEC (circles), in 1D at $a_s = 15 a_0$ (squares), and in 1D at $a_s = 819 a_0$ (diamonds). Here, the laser pulse length is fixed to $5$ms. The solid lines are exponentially damped sinusoids fit to the data. The dashed lines are linear fits in (a)-(c) and quadratic fits in (d) and (e) to the initial increase denoting the regime of linear response.}
\end{figure}

\subsection{The ABACUS algorithm}

The DSF for a homogeneous system is computed using the ABACUS algorithm. The computations are performed for a finite system of length $L$ with a finite particle number $N$ and with periodic boundary conditions. This means that the experimental situation is recovered only in the thermodynamic limit $N,L\rightarrow\infty$ with a density $n=N/L$ fixed through the local density approximation. We have confirmed that the particle numbers $N$ chosen for the computations (up to $N=128$) are large enough for the results to faithfully represent the thermodynamic limit. The only exceptions are the highest temperature curves in Fig.~3 (b)-(d) where some oscillations due to the finite size are still visible at high energies. The correlation function is evaluated numerically by summing contributions according to the Lehmann spectral representation. The spectral sum is infinite and needs to be truncated. The ABACUS algorithm performs this operation in an efficient way capturing the most relevant contributions. The error caused by the truncation is easily tractable by evaluating the $f$-sum rule and, for the presented data, does not exceed 5\% of the total spectral weight.

\subsection{Regime of linear response}

Measuring the DSF via Bragg spectroscopy requires to probe the system in the regime of linear response. This we tested experimentally by validating that the excitations created depend linearly on the pulse length and quadratically on the intensity of the Bragg lasers for the range of parameters used in the experiment, see Fig.~\ref{SuppFIG4} \cite{Brunello01MAT}.

\begin{figure}
\includegraphics[width=0.48\columnwidth]{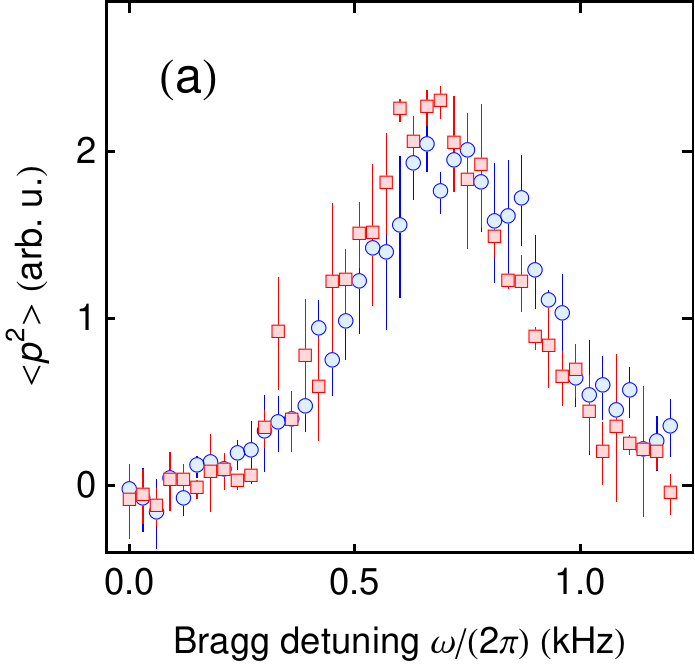}
\hspace{2mm}
\includegraphics[width=0.48\columnwidth]{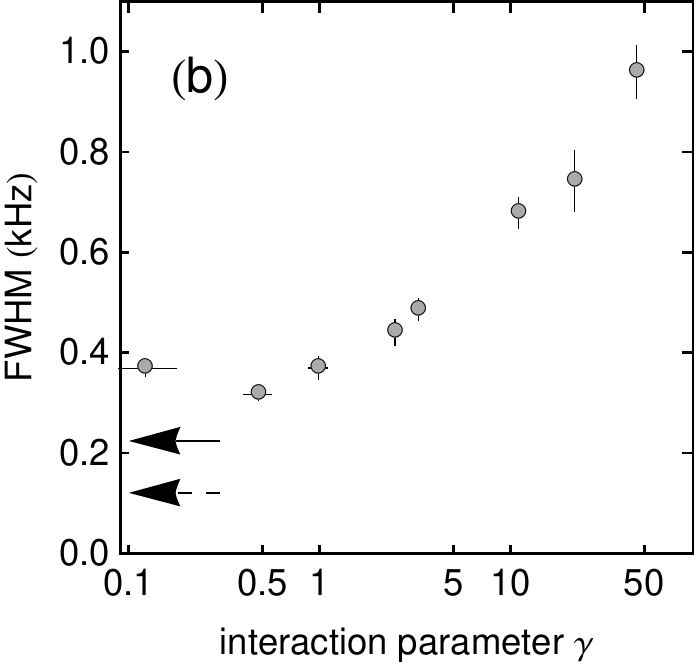}
\caption{\label{SuppFIG5} (a) Heating effects during the ramp of the scattering length $a_s$ on the Bragg spectroscopy data. Bragg spectroscopy data taken directly after the ramp of the scattering length to 173$a_0$ within 50 ms (squares) is compared to data taken after ramping to 819$a_0$ and back within 100 ms (circles). (b) Full width at half maximum (FWHM) extracted via the Gaussian fit (see main article) from the Bragg spectroscopy data as a function of $\gamma$. The dashed (solid) arrow denotes an estimate for line broadening due to the finite pulse length (quantum finite size effect) in the non-interacting limit.}
\end{figure}

\subsection{Heating effects on the excitation spectra}

Besides the detailed theoretical analysis of finite temperature effects on the measured DSF reported in the main article, we have checked in an experiment that heating during the ramp of $a_s$ to large values only marginally influences the shape of the excitation spectrum when compared to the effect of interactions. In Fig.~\ref{SuppFIG5}(a) we show Bragg excitation spectra taken at moderate interaction strength, corresponding to the data shown in Fig.~2(b) of the main article. For the two datasets shown, we either ramp directly to 173$a_0$ (squares) or ramp deep into the regime of strong interactions, $a_s$=810$a_0$, and back (circles) prior to applying the Bragg pulse.

\subsection{Interaction-independent spectral width at small $\gamma$}

In the limit of weak interactions we observe a width of the excitation spectra that levels off at a constant value as shown in Fig.~\ref{SuppFIG5}(b). We attribute this to the onset of two effects that prevent the observation of a $\delta$-like peak as expected in a non-trapped homogeneous system for  $\gamma \rightarrow 0$. First, the finite length of the Bragg excitation pulse causes Fourier broadening, which results in an estimated residual width of $\approx 120 \rm{Hz}$ in the non-interacting limit \cite{Pitaevskii03MAT}. Second, the finite size of the sample due to the presence of the harmonic trap leads to an uncertainty-limited energy width (not accounted for in the LDA) that can be estimated to $\delta \omega = 2 \hbar k / (m a_z )$ for a sample of non-interacting particles \cite{Golovach09MAT}. Here, $a_z = \sqrt{\hbar/(m \omega_{z})}$ denotes the quantum length scale of the trap.

\begin{figure}
\includegraphics[width=0.48\columnwidth]{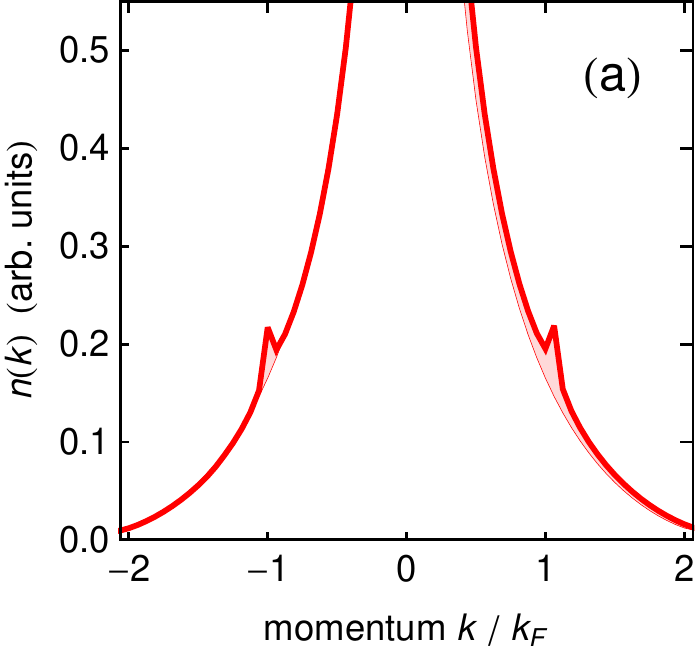}
\hspace{2mm}
\includegraphics[width=0.48\columnwidth]{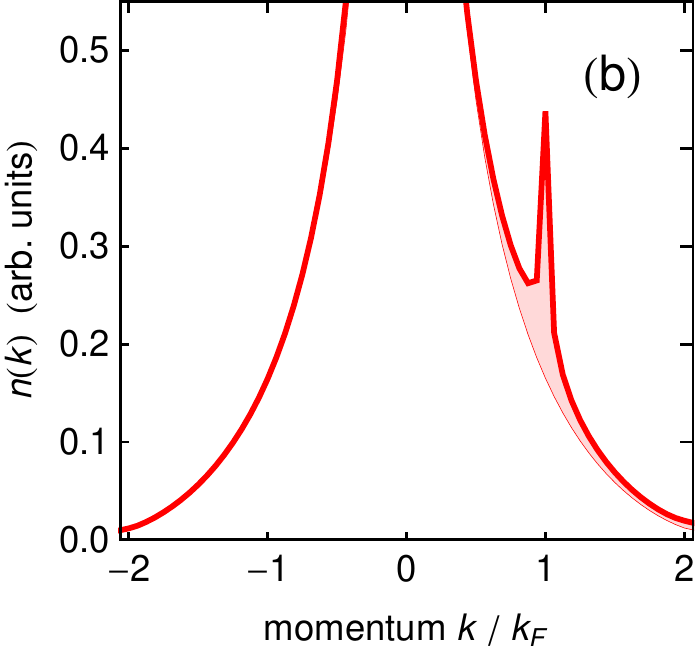}
\caption{\label{SuppFIG6} Momentum distribution of two states excited with a single (particle-hole) excitation over the ground state for $\gamma=64$. In the first case (a) the momentum is carried by a hole-like Lieb II excitation, in the second (b) by a particle-like Lieb I excitation. Both states have the same total momentum equal to $k/k_F=1$. The calculation is done for a system of 32 particles. The shaded areas indicate the difference to the ground-state momentum distribution.}
\end{figure}

\subsection{Momentum distribution of Lieb-I and Lieb-II excitations}

In order to illustrate the effects of additional particle-hole excitations on the momentum distribution, we consider two situations in which the total momentum of an excited state (obtained by adding a single particle-hole excitation with total momentum equal to the Fermi momentum on the ground state) is carried either by the hole (Lieb type II) mode (Fig.~\ref{SuppFIG6}(a)) or by the particle (Lieb type I) mode (Fig.~\ref{SuppFIG6}(b)). The computations are performed again using the ABACUS method~\cite{Caux07MAT}. The results show that while in both cases the momentum distribution function is augmented by broadened peaks, the hole excitation leads to an overall broadening of $n(k)$ while the particle excitation appears as a more distinguishable peak. This highlights the collective nature of the Lieb-II hole-like excitations and qualitatively reproduces the features seen in Fig.~4 of the main text.

\bibliographystyle{apsrev}

\begin{references}
\bibitem{calib}{We calibrate $k$ from the measured Bragg excitation spectrum of a weakly interacting BEC in 3D.}
\end{references}

\end{document}